\documentclass[journal=apchd5,manuscript=article, layout=twocolumn, email=true]{achemso}
\usepackage{chemformula} 
\usepackage[T1]{fontenc} 
\usepackage[utf8]{inputenc}

\usepackage{amsthm,amsmath,amssymb, amsfonts, mathtools}
\mathtoolsset{centercolon}
\usepackage{caption}
\usepackage{subcaption}
\usepackage{standalone}
\usepackage{tikz}
\usepackage{placeins}
\usetikzlibrary{shapes.geometric, arrows, positioning, calc}
\pgfdeclarelayer{bg}  
\pgfsetlayers{bg,main}
\usepackage[pdfencoding=auto]{hyperref}
\usepackage[capitalise, nameinlink]{cleveref}
\usepackage{nameref}
\usepackage{float}
\newcommand{\msgMatrix}[1]{\mathbf{#1}}
\newcommand{\msgVector}[1]{\mathbf{#1}}
\newcommand{\msgScalar}[1]{{#1}}
\newcommand{\scalarproduct}[2]{\left\langle #1, #2  \right\rangle}
\newcommand{\R}{ {\mathbb{R}} }
\newcommand{\SOne}{ {\mathbb{S}^1} }
\newcommand{\norm}[1]{\left\| #1 \right\|}

\newcommand{\transpose}[1]{#1^{\textnormal{T}}}

\newcommand{\varCylinder}{ {C} }

\newcommand{\varProjectionP}{\msgMatrix{P}_{\msgVector{u}\msgVector{v}}}
\newcommand{\varA}{\msgMatrix{A}_{|}}
\newcommand{\varShift}{{\msgVector{c}}}
\newcommand{\varInnerRange}{r^{-}}
\newcommand{\varOuterRange}{r^{+}} 
\newcommand{\varzPercent}{{z_{\%}}} 
\newcommand{\varx}{\mathtt{x}}
\newcommand{\vary}{\mathtt{y}}
\newcommand{\varz}{\mathtt{z}}
\newcommand{\varEFOV}{{\digamma}}
\newcommand{\varpEFOVC}[1]{{\digamma}^{+}_{#1}}
\newcommand{\varnEFOVC}[1]{{\digamma}^{-}_{#1}}
\newcommand{\varpEFOVX}{{\varpEFOVC{x}}}
\newcommand{\varnEFOVX}{{\varnEFOVC{x}}}
\newcommand{\varpEFOVY}{{\varpEFOVC{y}}}
\newcommand{\varnEFOVY}{{\varnEFOVC{y}}}
\newcommand{\varpEFOVZ}{{\varpEFOVC{z}}}
\newcommand{\varnEFOVZ}{{\varnEFOVC{z}}}
\newcommand{\varDataset}{{\mathbb{D}}}

\newcommand{\image}{{\msgMatrix{I}}}
\newcommand{\Nz}{{N_{\Delta z}}}
\newcommand{\NperZ}{N_z}
\newcommand{\threshold}{{\tau}}
\newcommand{\Nmax}{{n_p}} 
\newcommand{\Nfreqs}{{K}}

\author{Manuel Kunisch}
\affiliation[equal]{these authors contributed equally}
\affiliation[Bielefeld University]{Department of Physics, Bielefeld University, 33615 Bielefeld, Germany}
\email{mkunisch@physik.uni-bielefeld.de} 
\author{Sascha Beutler}
\affiliation[equal]{these authors contributed equally}
\affiliation[University of Münster]{Institute for Computational and Applied Mathematics, University of Münster, 48149 Münster, Germany}
\email{sascha.beutler@uni-muenster.de}
\author{Christian Pilger}
\affiliation[Bielefeld University]{Department of Physics, Bielefeld University, 33615 Bielefeld, Germany}
\author{Friedemann Kiefer}
\affiliation[University of Münster]{European Institute for Molecular Imaging, University of Münster, 48149 Münster, Germany}
\author{Thomas Huser}
\affiliation[Bielefeld University]{Department of Physics, Bielefeld University, 33615 Bielefeld, Germany}
\author{Benedikt Wirth}
\affiliation[University of Münster]{Institute for Computational and Applied Mathematics, University of Münster, 48149 Münster, Germany}

\title
  {Active Axial Motion Compensation in Multiphoton-Excited Fluorescence Microscopy}

\abbreviations{IR,NMR,UV}

\SectionNumbersOff 
\usepackage{afterpage}

\begin{document}
	
\begin{abstract}
    In living organisms, the natural motion caused by the heartbeat, breathing, or muscle movements leads to the deformation of tissue caused by translation and stretching of the tissue structure. This effect results in the displacement or deformation of the plane of observation for intravital microscopy and causes motion-induced aberrations of the resulting image data. This, in turn, places severe limitations on the time during which specific events can be observed in intravital imaging experiments. These limitations can be overcome if the tissue motion can be compensated such that the plane of observation remains steady. We have developed a mathematical shape space model that can predict the periodic motion of a cylindrical tissue phantom resembling blood vessels. This model is then used to rapidly calculate the future position of the plane of observation of a confocal multiphoton fluorescence microscope. The focal plane is continuously adjusted to the calculated position with a piezo-actuated objective lens holder. We demonstrate active motion compensation for non-harmonic axial displacements of the vessel phantom with a field of view up to 400 µm $\times$ 400 µm, vertical amplitudes of more than 100 µm, and at a rate of 0.5 Hz.
\end{abstract}

\section{Introduction}
Imaging physiological events deep within living organisms with the highest spatial resolution typically requires the specificity of fluorescence microscopy. Currently, the most common way of accomplishing these challenging imaging conditions (commonly known as "intravital microscopy") is realized by two- and three-photon fluorescence excitation with femtosecond laser pulses \cite{Scheele_2022}. In inflammation research, this type of intravital microscopy has, for example, enabled the in vivo imaging of leukocyte migration dynamics with unprecedented spatial resolution and on very short time scales \cite{mcardle_intravital_2015,ji_technologies_2016}. Two-photon (2P) fluorescence excitation allows for deep tissue imaging with excellent axial sectioning capability due to reduced light scattering, higher penetration depth, and smaller excitation volumes \cite{ji_adaptive_2017}. Imaging neural activity through a cranial window in mechanically fixed mice can typically be achieved without the image quality being affected by major motion artifacts. Imaging experiments with single cell resolution, e.g. in the thorax, however, are limited by physiological tissue movement caused by the respiratory and cardiovascular system \cite{Bakalar_2012,Vinegoni_2014}. In mice, this leads to an axial displacement of the plane of observation of up to 1,200 $\mu$m every second due to chest movement during breathing and by up to 50 $\mu$m with a rate of 300-600 beats per minute caused by the heartbeat. Various methods have been developed for dealing with and compensating this motion, for example by mechanical restriction of the motion \cite{Lee_2012}, or by time-gated imaging, where the acquisition is either triggered by or images are discarded based on vital parameters measured simultaneously during the imaging process by electrocardiography \cite{Vinegoni_2012,mcardle_intravital_2015}. All of these methods do, however, carry certain disadvantages. The hard mechanical stabilization of organs can, for example, cause damage to the tissue or induce additional inflammatory triggers that will likely interfere with the biological outcome, in particular when imaging immune cell response in vivo \cite{Vinegoni_2014}. Vital parameter-triggered imaging followed by post-recording data selection, on the other hand, will result in a substantial amount of previously acquired data having to be discarded due to poor registration. Thus, the limits in spatial resolution and image quality can only be pushed further by actively improving the image acquisition process itself in order to avoid image-degrading motion artefacts.

\begin{figure*}[ht]
    \centering
    \includegraphics[width=\textwidth]{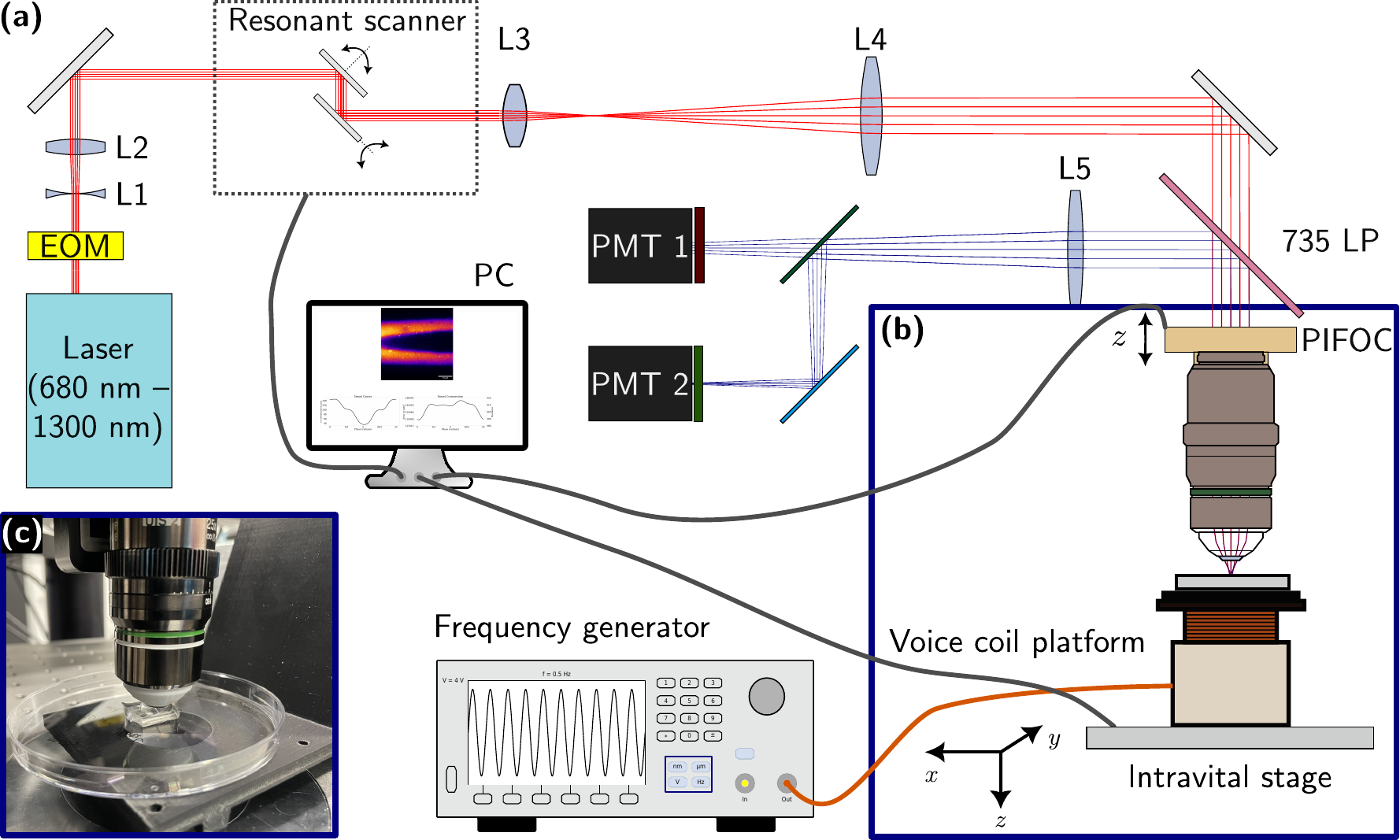}
    \caption{\textbf{Optical setup and simulation of vital motion} (a) The optical setup consists of a tunable femtosecond laser source, which is rapidly shuttered by an electrooptic modulator (EOM) and directed into the scanning 2P-microscope via a telescope (consisting of lenses L1 and L2) . Two resonant galvanometric scanning mirrors raster-scan the laser beam, which is directed to the backfocal plane of the objective lens by a second telescope (L3 and L4). The excitation light is focussed into the sample,  which is held on a manually adjustable $xyz$-stage.
    The fluorescence emission is detected in epi-direction, separated from the excitation light by a 735 nm long-pass filter, and forwarded to the photomultiplier tubes (PMTs) via a tube lens (L5). A combination of bandpass and long/shortpass filters are used to solely detect the desired signal emitted by the sample.
    (b) A voice coil platform connected to a frequency generator is used to simulate the $z$-movement. The frequency and amplitude of the desired sample motion can be adjusted to actively control the sample position with regard to the microscope's focal plane. The data collected from the measurement (see \cref{fig:PhantomComparison}) is evaluated by a custom-written algorithm, which in turn controls the piezo-stage of the objective lens holder (PIFOC). (c) Photograph of the polydimethylsiloxane (PDMS) sample phantom held in a Petri dish underneath the objective lens.
    }
    \label{fig:Setup}
\end{figure*}

An alternative way to overcome the fast motion-induced axial shifts of the image plane is to actively align the focal plane with the movement of the sample by continuously adjusting either the position of the objective lens or the sample stage in a confocal laser scanning microscope. This was demonstrated by Bakalar et al., who used fast 3D cross-correlation of sequentially acquired image volumes to detect displacements and adjust the tissue position to compensate for sample drift on the second to minute time scale \cite{Bakalar_2012}. Here, we demonstrate fast motion compensation by actively aligning the focal plane for movements on the sub-second time scale for samples movements of up to 200 µm and fields of view of 400 µm $\times$ 400 µm. A tissue phantom resembling the dimensions of murine carotid arteries is actuated by regular, non-harmonic axial movement. Images where the sample moves through the focal plane are collected, processed, and the sample movement along the axial direction is then extracted. The extraction is based on a mathematical model, which uses algebraic surfaces to describe the general shape of blood vessels, here carotid arteries, and fits the surface with a linear regression in a spirit similar to static surface fitting \cite{Vaughan87,Juettler00,BertAlf02}. 
Linear regression with a surface model tailored to vessels (rather than with more flexible, but unspecific kernel methods as often used in imaging \cite{Steidl2015}) has the advantage that motion parameters of the full vessel can more easily be extracted.
Feeding a signal derived from the mathematical model to a piezo-electric displacement stage, which moves the objective lens, allows us to actively follow and compensate the fast axial sample movement in order to maintain the focal plane (see \cref{fig:Setup}). This approach is versatile and could also be extended and applied to compensate distortions of the vessel shape e.g. during peristaltic blood flow.

\section{Methods}
\afterpage{\FloatBarrier}

\subsection{Vessel Phantom}
\label{sec:FabricationPhantom}
In order to mimic blood vessels of living organisms a phantom is required that resembles the vessels in terms of their dimensions, elastic behavior, and their ability to be stained with fluorophores. The diameter of the carotid artery is on the order of 400 µm  \cite{Sutton2008}, which makes it somewhat challenging to produce a phantom of this structure. Most commercially available options, such as flexible fluid-carrying tubes typically have larger diameters. Furthermore, it is crucial for the imaging process that the replica exhibits staining similar to its realistic counterpart. This involves either surface functionalization of the phantom or employing additional techniques, such as post-processing of images to selectively capture fluorescent signals only from the phantom's inner boundary.
Ultimately, we decided to pursue a custom-made solution based on the organic polymer polydimethylsiloxane (PDMS). Depending on the amount of curing agent, PDMS tissue phantoms can be made with varying degrees of elasticity and they are widely used in microfluidic lab-on-a-chip applications to support the realization of artificial organs (organoids) \cite{Nge2013Apr} \cite{Saorin2023Jul} .
Tissue phantoms can easily be created with commercially available PDMS kits consisting of the polymer base and a curing agent. For the fabrication of our phantom, PDMS was purchased from Mavom (Sylgard 184).
This approach offers several benefits. The main advantage is the flexible choice of elasticity controlled by the ratio of the polymer base to the curing agent. Choosing a ratio of 1:100 of curing agent to polymer base resulted in a solid and rather stiff block of PDMS after the curing process. We found this to be ideal for our experiments as the current study is focused exclusively on the compensation of respiratory-like sample motion with the help of a novel mathematical model. In order to create a small, liquid-carrying vessel within the PDMS block, a thin electric wire used in loudspeaker coils was embedded in the PDMS during the curing process. 

The choice of a coil wire had several benefits: coil wires are very tear-resistant so they can be easily removed from the PDMS even well after the end of the curing process. This results in a cylindrical cavity. Furthermore, coil wires are available with outer diameters ranging from a few 10s to several 100s of microns and, thus, can be chosen to perfectly match the size of a murine carotid artery (see \cref{fig:Phantom}).
This provides added flexibility for the creation of even smaller or larger vessel phantoms in the future.
Here, a coil wire with an outer diameter of $\sim $ 370 µm was cut into short pieces of $\sim 10~\text{cm}$ length and inserted into a container filled with liquid PDMS. The PDMS was then left to cure for 60 min in an oven at $60\,^\circ$C. The coil wire was subsequently carefully pulled out of the hardened PDMS block with pliers. This resulted in a narrow, hollow channel in the otherwise solid PDMS.

A thin slice of the PDMS block containing such a channel was then cut out of the polymer with a surgical scalpel. In preparation for 2P-excited fluorescence imaging, a liquid solution of the dye Nile Blue using water as a solvent  (concentration $0.05~\text{mg/mL}$ diluted with water, emission peak at $\sim 660~\text{nm}$) was then added to the channel with the help of a syringe.

\subsection{2P Fluorescence Imaging of the Phantom} 
 
 \begin{figure*}[ht]
 	\centering
 	\includegraphics[width=\textwidth]{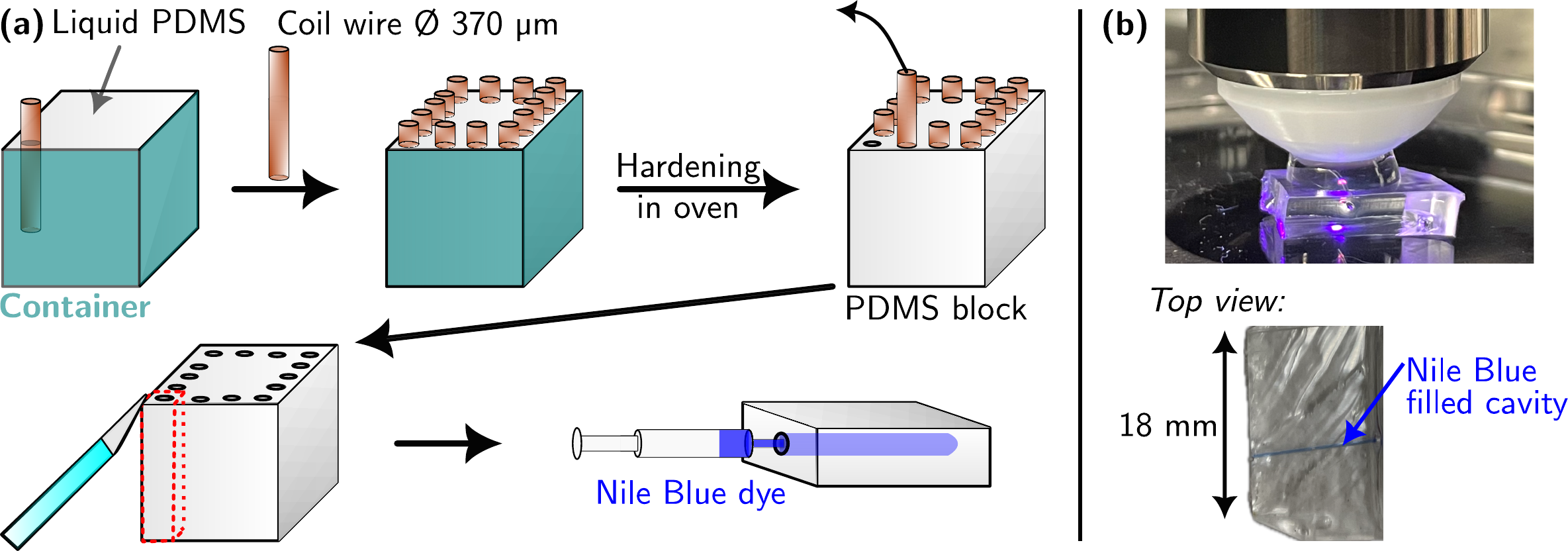}
 	\caption{\textbf{Manufacturing process and imaging of the PDMS vessel phantom} -- (a) A thin electrical coil wire is inserted into a container with liquid PDMS and cured in an oven at $60\,^\circ\text{C}$ for 60 minutes. After hardening, the coil wire is carefully pulled out, leaving a hollow channel in the solid PDMS block. A single channel is cut out of the block and stained with a liquid solution of Nile Blue dye dissolved in doubly-distilled water.  (b) A PDMS slice with dimensions of approx.\ 2 cm $\times$ 1 cm $\times$ 0.5 cm containing a hollow channel is used for imaging by 2-photon fluorescence excitation underneath the 25x objective lens. Widefield LED illumination of the cavity filled with the fluorescent dye results in fluorescence emission from a large area as can be seen by the naked eye in the photograph. 
 		The individual fluorescence images acquired at different depths within the sample represent axial planes within the cylindrical cross-section of the phantom.
 	}
 	\label{fig:Phantom}
 \end{figure*}
 
 \begin{figure*}[h]
 	\centering
 	\includegraphics[width=\textwidth]{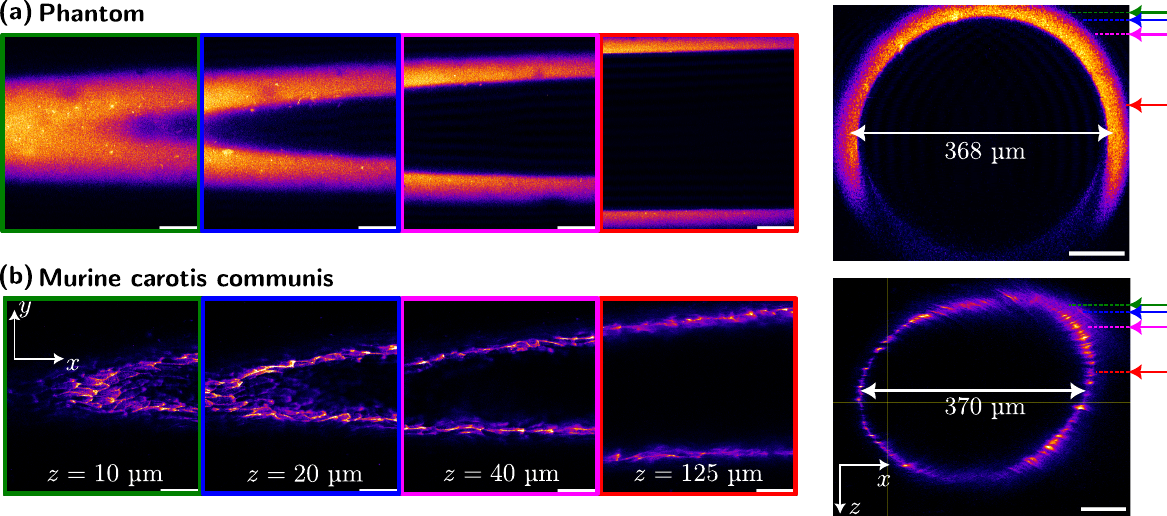}
 	\caption{\textbf{Comparison of the resting vessel phantom with a fluorescently stained murine carotis communis sinistra} -- (a) Four representative 2-photon excited fluorescence image frames acquired by translating the objective lens to obtain a full volumetric scan of the stationary vessel phantom (see \cref{fig:Setup}). Additionally, the $xz$ cross-section of the phantom is visualized from the reconstruction of more than 350 such frames, spaced 1 µm apart. The colored arrows indicate the location of the $z$-planes of the individual frames shown in (a). The $xz$ cross-section indicates an inner diameter of the phantom of 368 µm based on the image data. (b) Selected images from a volumetric ($z$-series) scan acquired during 2P fluorescence image acquisition of a murine carotid artery and its orthogonal $xz$ perspective. Here, the endothelial cells of the vessel were stained by immunofluorescence staining to achieve the vessel contrast. Based on the evaluation of the cross-section the diameter of this carotid artery is 370 µm, which corresponds quite well to the inner dimension of our vessel phantom.
 		The images of the phantom reveal that the hydrophobic interaction of the dye-water solution with the PDMS block leads to similar contrast as that obtained by staining the surface of endothelial cells within an artery.
 		Scale bar: 75 µm.}
 	\label{fig:PhantomComparison}
 \end{figure*}
The phantom was subsequently imaged by 2P fluorescence excitation on a TriMScope Matrix (Miltenyi Biotec, Bielefeld, Germany, see \cref{fig:Setup}). Two-photon excitation was achieved with a tunable fs-laser system (Insight X3, Spectra-Physics). The size of the output beam was expanded with a lens telescope (composed of lenses L1 and L2) to fill the back focal plane of the microscope objective lens. A pair of resonant galvanometric scanning mirrors was used to raster-scan the laser beam and to enable image acquisition with frame rates of 30 frames per second (fps) or higher. Such imaging rates are typically used for the study of living animal models, such as mice \cite{Arasa_2021} \cite{Martens_2020} . The pivot point of the scanning mirrors was projected to the back focal plane of the objective lens by the scanning telescope (L3 and L4). 
All data were acquired using a 25x 1.05 NA water immersion objective lens (Olympus XLPLN25XWMP2, Evident Europe GmbH, Hamburg, Germany), which is optimized for infrared excitation wavelengths. The large working distance of 2 mm enables deep tissue imaging for in vivo experiments. Moreover, the 25x objective lens also offers a large field of view of approx. 400 µm $\times$ 400 µm, which is essential for capturing the entire cross-section of the carotid artery phantom as shown in \cref{fig:Phantom}. 
The specimen is held on a $xyz$-stage, while being raster-scanned by 2P fluorescence excitation in a point by point fashion using an excitation wavelength of 1100 nm. The fluorescence signal was collected by epi-detection and separated from background signals by a combination of longpass (LP 735nm) and bandpass (690/50) filters and detected by photomultiplier tubes (PMTs) via an additional lens (L5) (see \cref{fig:Setup}).

The excellent $z$-sectioning ability of multi-photon fluorescence excitation (which is confined to the focus spot) allows one to resolve the focal plane with an axial resolution of $\simeq 1$ µm. This capability is exploited for the reconstruction and evaluation of imaging data of the phantom. 

The simulation of the movement induced by a mouse's heart beat or breathing, was carried out by adapting a voice coil platform to the multi-photon-excitation fluorescence microscope (see \cref{fig:Setup} (b)). This voice coil platform was built out of a subwoofer driver. A lightweight, 3D-printed mounting platform was glued to the central actuator of the subwoofer. By connecting the loudspeaker to a frequency generator (Hameg, HMF2550), high-precision tuning of the supply voltage and frequency was accomplished, which allowed us to dynamically alter the focal plane shift of the sample. Thus, we could tune and simulate the sample motion induced by the respiration. The data collected during a measurement run (see \cref{fig:PhantomComparison} (a)) were evaluated by a custom-written algorithm based on the expected shape of the object to be tracked (see \cref{fig:MathematicalFlowChart}), which, in turn, controlled the piezo-actuated objective lens holder (PIFOCP-725.xCDE2, Physik Instrumente (PI) GmbH \& Co. KG, Karlsruhe, Germany) and was, thus, able to correct the induced movement.

\subsection{Mathematical Shape Model} \label{subsection:MathematicalModel}
\label{sec:MathematicalModel}
Typical algorithms for determining movements rely on image sequences containing 3-dimensional information such as photos of 3D scenes \cite{Oron18,Hur21}.
In our setting, the individual microscopy images only provide 2-dimensional information; 3-dimensional information has to be assembled from these 2D images over time.
Therefore, our mathematical motion prediction and compensation model consists of two phases.
In the \emph{motion analysis phase} the model learns the (periodic) vessel deformation from information obtained from multiple images,
and in the \emph{motion correction phase}, it stabilizes the focal plane relative to the vessel surface using the actuated optics to compensate for the temporal motion.
We start by introducing a few basic concepts and our notation.

\paragraph{Truncated Fourier series.} \label{paragraph:TruncatedFourierSeries}
We will represent the temporal dependence of the deformation parameters
by a truncated Fourier series with truncation frequency $\Nfreqs \in \mathbb{N}$,
i.e.\ by an expression of the form
\begin{equation*}
	\sum_{k=0}^{\Nfreqs-1} f_{k+1} \cos(2 \pi k \theta) 
	+ \sum_{k=1}^{\Nfreqs-1} f_{(k + \Nfreqs)} \sin(2 \pi k \theta) 
	,
\end{equation*}
with real or vector-valued coefficients $f_i$, $i \in \{1, \ldots, 2 \Nfreqs -1\}$,
where $\theta\in\R$ represents time or rather the phase of the periodic deformation.

\paragraph{Barycenter fitting.} \label{paragraph:Barycenter}
Given a data set $\left(x_i, y_i, z_i, \theta_i\right)_{i}\subset\R^4$ of points in spacetime (such as a list of bright voxels) 
we approximate the temporally varying barycenter $\varShift:\R\to\R^3$ as \emph{that} \hyperref[paragraph:TruncatedFourierSeries]{truncated Fourier series} whose coefficients are the minimum-norm solution of $\varShift(\theta_i) = (x_i, y_i, z_i)$ for all $i$.

\paragraph{Cylinder.} \label{paragraph:Cylinder}
The imaged blood vessel is approximated by a cylinder
$$
\varCylinder :=	\left\{ \msgMatrix{A} \msgVector{x} + \msgVector{d} + \msgScalar{s} \msgVector{w} : \msgVector{x} \in \SOne, \ \msgScalar{s} \in \mathbb{R} \right\}
$$
of elliptical cross-section,
in which the unit vector $\msgVector{w}\in\R^3$ indicates the axis, $\msgVector{d}$ is an offset, and $\msgMatrix{A} \in \R^{3 \times 2}$.
Extending $\msgVector{w}$ to an orthonormal basis $\{\msgVector{u},\msgVector{v},\msgVector{w}\}$ and abbreviating $\varProjectionP=[\msgVector{u}\ \msgVector{v}] \in \R^{3 \times 2}$,
we can take $\msgMatrix{A} = \varProjectionP \varA$ for a regular matrix $\varA \in \R^{2 \times 2}$ (otherwise the cylinder would be degenerate). 
Given a point $\msgVector{p} \in \R^3$, its cylinder coordinates $(\msgVector{x},s)\in\R^2\times\R$ are
\begin{equation*}
	s = \scalarproduct{\msgVector{p} - \msgVector{d}}{\msgVector{w}},\quad 
	\msgVector{x} = \varA^{-1} \transpose{\varProjectionP} \left( \msgVector{p} - \msgVector{d} \right),
\end{equation*}
i.e.\ $\msgVector{x}$, $s$ and $\msgVector{p}$ are related by $\msgVector{p} = \msgMatrix{A} \msgVector{x} + s \msgVector{w} + \msgVector{d}$.

\paragraph{Cylinder construction.} \label{paragraph:CylinderConstruction}
We approximate the blood vessel shape by
fitting an algebraic surface of degree $2$ to the fluorescently labeled inner boundary of the vessel phantom, i.e.\ a surface of the form
\begin{align*} 
	\left\{ \msgVector{p} \in \R^3 : \transpose{\msgVector{p}} \msgMatrix{M} \msgVector{p} +  \transpose{\msgVector{b}} \msgVector{p}  = 1 \right\}
\end{align*}
with $\msgMatrix{M}\in\R^{3\times3}$ symmetric and $\msgVector{b}\in\R^3$.
This also allows a deviation from a perfect cylinder,
e.g.\ an ellipsoidal or hyperboloidal shape.
For further processing, however, we sometimes ignore this deviation
and construct the closest cylinder as follows:
Let $\{\msgVector u,\msgVector v,\msgVector w\}$ denote the unit eigenvectors of $\msgMatrix M$
with eigenvalues $\mu_1,\mu_2>0$ and a third eigenvalue $\mu_3\in\R$ close to zero
(since the surface approximates a cylinder).
The cylinder axis is thus $\msgVector{w}$, and the cylinder is obtained by modding out variations along this direction via the projection $\msgMatrix{P}=\varProjectionP\transpose{\varProjectionP}$,
\begin{multline*} 
	\left\{ \msgVector{p} \in \R^3 : \transpose{\msgVector{p}} \msgMatrix{P}\msgMatrix{M}\msgMatrix{P} \msgVector{p} +  \transpose{\msgVector{b}}\msgMatrix{P}\msgVector{p}  = 1 \right\}\\
	=\left\{ \msgVector{p} \in \R^3 : \transpose{(\msgVector{p}-\msgVector{d})}\msgMatrix{C}(\msgVector{p}-\msgVector{d}) = 1 \right\},
\end{multline*}
where $\msgVector{d}\in\R^3$ solves $-2\msgMatrix{M}\msgVector{d}=\msgMatrix{P}\msgVector{b}$
and $\msgMatrix{C}=\msgMatrix{M}/(1+\transpose{\msgVector{d}}\msgMatrix{M}\msgVector{d})$.
The cylinder parameters then turn out to be the axis $\msgVector{w}$, offset $\msgVector{d}$, and
\begin{align*}
	\msgMatrix{A} := \varProjectionP \varA
	\text{ with }
	\varA := \mathrm{diag}(\tfrac1{\sqrt{\lambda_1}},\tfrac1{\sqrt{\lambda_2}})
\end{align*}
for $\lambda_i=\mu_i/(1+\transpose{\msgVector{d}}\msgMatrix{M}\msgVector{d})$ the eigenvalues of $\msgMatrix{C}$.

\paragraph{Coordinates.}
The field of view is denoted $$\varEFOV := [\varnEFOVX, \varpEFOVX] \times [\varnEFOVY, \varpEFOVY] \times [\varnEFOVZ, \varpEFOVZ].$$
We define $\Nz \in \mathbb{N}$ equidistant heights
\begin{align*}
	\varz_i &:= \varnEFOVZ + (i-1) \ \frac{\varpEFOVZ-\varnEFOVZ}{\Nz-1},
	\quad i = 1,\ldots,\Nz,
\end{align*}
at which we record two-dimensional images $\image \in \R^{n \times m}$.
We assign to the pixels of $\image$ the coordinates 
\begin{align*}
	\varx_i &:= \varnEFOVX + \frac{2i-1}{2n} (\varpEFOVX - \varnEFOVX),\quad i = 1, \ldots, n \\
	\vary_j &:= \varnEFOVY + \frac{2j-1}{2m} (\varpEFOVY - \varnEFOVY),\quad j = 1, \ldots, m.
\end{align*}

\paragraph{Motion analysis phase.} \label{paragraph:MotionAnalysisPhase}
We assume a periodic motion (of known frequency for the current publication) so that each time $t$ can be associated with a phase $\theta(t) \in [0, 2 \pi)$. 
At each height $\varz_r$ we acquire $\NperZ$ images (which all differ from each other due to the motion),
so $\Nz \NperZ$ images in total.
Let the acquisition of the $k$th image $\image^k\in\R^{n\times m}$ start at time $t_k$ and let $\varz_{r_k}$ be the corresponding height,
then we save the data set $\varDataset := \bigcup_{k=1}^{\Nz \NperZ} \varDataset_k$ with
\begin{align*}
	\varDataset_k := \left\{ (\varx_i, \vary_j, \varz_{r_k}, \theta(t_k)): \image^k_{i,j} \ge \tau \right\},
\end{align*}
of pixels brighter than $\tau$ (those will lie in the simulated "endothelium" of the tissue phantom).
For a smaller memory footprint we reduce $\varDataset$ further to a data set $\widetilde{\varDataset}\subset\varDataset$ of just $\Nmax$ elements.
We fit a periodically changing algebraic surface of degree $2$ to $\widetilde{\varDataset}$. 
At a fixed phase $\theta$ it is described by
\begin{align*} 
	\left\{ \msgVector{p} \in \R^3 : \transpose{\msgVector{p}} \msgMatrix{M}(\theta) \msgVector{p} +  \transpose{\msgVector{b}(\theta)} \msgVector{p}  = 1 \right\},
\end{align*}
where now $\msgMatrix{M}(\theta) \in \R^{3 \times 3}$ and $\msgVector{b}(\theta) \in \R^3$ depend on the phase.
By using the one-level set we fix a particular scaling of $\msgMatrix{M}(\theta)$ and $\msgVector{b}(\theta)$.
However, to improve the stability of fitting $\msgMatrix{M}(\theta)$ and $\msgVector{b}(\theta)$
(note, e.g., that the zero vector cannot fulfill the equation),
we first shift the coordinate origin to the \hyperref[paragraph:Barycenter]{barycenter}, which is equivalent to replacing $\widetilde{\varDataset}$ with
\begin{align*}
	\widehat{\varDataset} := \left\{ \left((x,y,z) - \varShift(\theta), \theta\right) : \left(x,y,z,\theta\right)\in \widetilde{\varDataset} \right\}.
\end{align*}
Now we just describe $\msgMatrix{M}(\theta)$ and $\msgVector{b}(\theta)$ as truncated Fourier series
and determine their coefficients as the least squares solution of
\begin{align}\label{eq:algebraicSurface}
	\transpose{\msgVector{p}} \msgMatrix{M}(\theta_{\msgVector{p}}) \msgVector{p} +  \transpose{\msgVector{b}(\theta_{\msgVector{p}})} \msgVector{p}  = 1
	\quad\text{for all }(\msgVector{p}, \theta_{\msgVector{p}}) \in \widehat{\varDataset}.
\end{align}
Via the previous \hyperref[paragraph:CylinderConstruction]{cylinder construction}
we can then turn $(\msgMatrix{M}(\theta), \ \msgVector{b}(\theta))$
into cylinder parameters $\left(\msgMatrix{A} = \varProjectionP \varA, \  \msgVector{d}, \ \msgVector{w} \right)$,
where for notational simplicity we suppressed their explicit dependence on $\theta$
and where the coordinate shift from $\widetilde{\varDataset}$ to $\widehat{\varDataset}$ can be undone by adding $\varShift(\theta)$ to $\msgVector{d}$.

\begin{figure*}[h!]
	\centering
	\includegraphics[width=\textwidth]{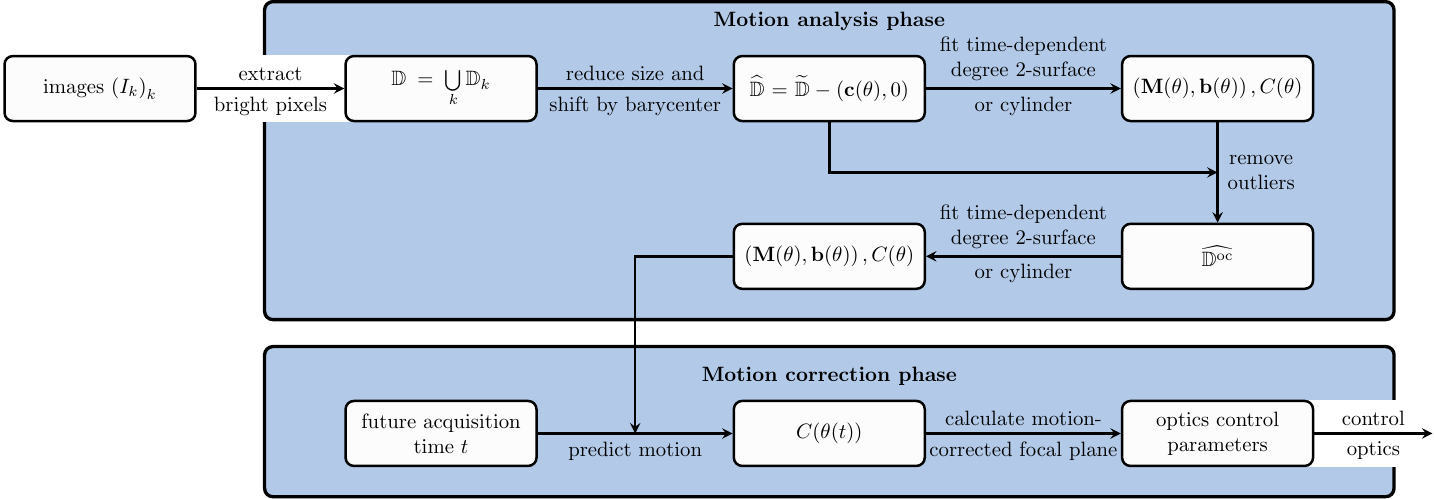}
	\caption{Flow chart of the motion correction algorithm.
		$\varDataset_k$ are the spacetime coordinates of all bright pixels in image $I_k$.
		$\widetilde{\varDataset}$ is a subset of $\varDataset$ and $\varShift(\theta)$ the temporally changing barycenter of $\widetilde\varDataset$.
		$\widehat{\varDataset^{\text{oc}}}$ is the outlier corrected and barycenter-shifted data set. 
		$\msgMatrix{M}$ and $\msgVector{b}$ are parameters defining a degree-2 surface,
		$\varCylinder$ is a cylinder.
		All fitted quantities depending on $\theta$ are represented via a truncated Fourier series.
	}
	\label{fig:MathematicalFlowChart}
\end{figure*}

Based on this first approximation of a "deforming blood vessel" by a temporally changing cylindrical shape $\varCylinder(\theta)$,
we now correct for potential outliers in $\widetilde{\varDataset}$:
We represent each point $\msgVector{p}$ with $(\msgVector{p}, \theta) \in \widetilde{\varDataset}$ in \hyperref[paragraph:Cylinder]{cylinder coordinates} $(\msgVector{x},s) \in \R^2 \times \R$ for $\varCylinder(\theta)$.
Then $\msgVector{p}\in \varCylinder(\theta)$ if and only if $\norm{\msgVector{x}} = 1$.
Therefore we remove all $(\msgVector{p}, \theta)$ from $\widetilde{\varDataset}$ that violate
\begin{align*} 
	\varInnerRange \le \norm{\msgVector{x}} \le \varOuterRange
\end{align*}
for some fixed $\varInnerRange < 1 < \varOuterRange$.
We denote the outlier-corrected data set by $\varDataset^{\text{oc}}\subset\widetilde\varDataset$
and the barycenter-shifted dataset by $\widehat{\varDataset^{\text{oc}}}$.
Replacing $\widehat\varDataset$ with $\widehat{\varDataset^{\text{oc}}}$ and following the same steps as before, we get an outlier-corrected temporally changing degree-2 surface $(\msgMatrix{M}(\theta), \ \msgVector{b}(\theta))$ or cylinder $\varCylinder(\theta)$.
The overall procedure is summarized in \cref{fig:MathematicalFlowChart}.

\paragraph{Motion correction phase.} \label{paragraph:MotionCompensationPhase}
Multiple motion compensation objectives can be imagined.
For instance, one could have the optics follow a fixed chosen point on the endothelium
whose motion can be stabilized based on the results of the \hyperref[paragraph:MotionAnalysisPhase]{motion analysis phase}.
Instead, here we restrict ourselves to describing how to stably keep the focal plane at a prescribed relative z-position of the imaged vessel structure,
e.g.\ $\varzPercent=40\,\%$ of the (potentially changing) vertical vessel extension above the (moving) vessel center.

For each image to be acquired we choose an acquisition time $t$ sufficiently in the future
to still allow the computation and the controlled approach of the motion-corrected $z$-position.
To find the vertical extension of $\varCylinder(\theta(t))=\left\{ \msgMatrix{A} \msgVector{x} + \msgVector{d} + \msgScalar{s} \msgVector{w} : \msgVector{x} \in \SOne, \ \msgScalar{s} \in \mathbb{R} \right\}$ at the vessel center, we maximize and minimize
$
\langle\msgMatrix{A} \msgVector{x},\msgVector{e_3}\rangle
$
over $\msgVector{x}\in\SOne$, yielding $\msgVector{x}_{\substack{\max\\\min}}=\pm\transpose{\msgMatrix{A}}\msgVector{e_3}/\norm{\transpose{\msgMatrix{A}}\msgVector{e_3}}$.
The vertical extension thus is $\langle\msgMatrix{A}(\msgVector{x}_{\max}-\msgVector{x}_{\min}),\msgVector{e_3}\rangle$,
and the targeted focal plane therefore has the $z$-coordinate
\begin{align*}
	\langle -\varzPercent\msgMatrix{A}(\msgVector{x}_{\max}-\msgVector{x}_{\min})+\msgVector{d},\msgVector{e_3}\rangle.
\end{align*}
In our experiment we used the following parameters: 
\begin{center}
	\begin{tabular}{ c | c | c | c | c | c | c | c}
		$\Nz$ &  $\NperZ$ & $\threshold$ & $\Nmax$ 		& $\Nfreqs$ & $\varzPercent$ & $\varInnerRange$ & $\varOuterRange$ \\ 
		\hline
		$100$ &  $30$ 	  & $900$		 & $1.5e7$ 	&	$6$		& $0.425$		& 	$0.8$			& $1.1$  
	\end{tabular}
\end{center} 

\section{Results and Discussion}
\begin{figure*}[ht]
	\includegraphics[width = \textwidth]{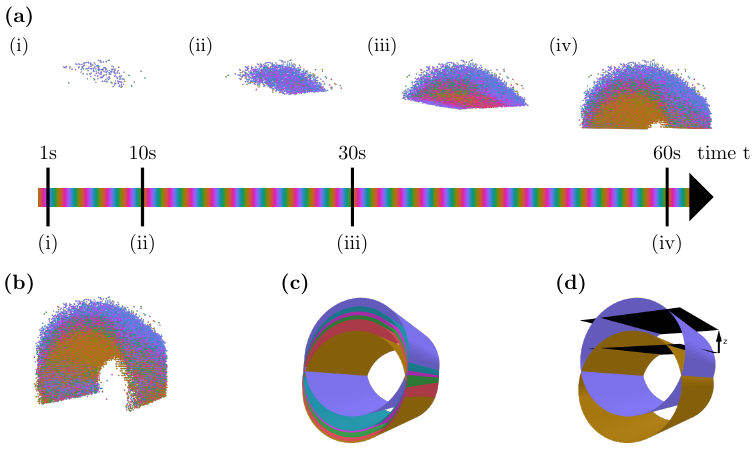}
	\caption{(a): A timeline where every timepoint is colored by the corresponding phase of the motion. We used a motion of $0.5$ Hz and therefore the colors repeat every 2 seconds. At some timepoints a marker labed with the subfigure name of the corresponding pointcloud is shown which is acquired until then.
		(i)-(iv): pointcloud of the collected datapoints. Each point is colored by the phase of its acquisition.
		(b): pointcloud of the datapoints after the acquisition of the last image in the \hyperref[paragraph:MotionAnalysisPhase]{motion analysis phase}. 
		(c): a visualization of cylinders $\varCylinder(\theta)$ at 5 different phases $\theta$. 
		(d): schematic illustration of the \hyperref[paragraph:MotionCompensationPhase]{motion correction phase}. Two cylinders at different phases and the adjustment of the focal plane that is required such that the $z$-position relative to the cylinder is the same. }
	\label{fig:MathematicalTimeLine} 
\end{figure*}
\subsection{Comparison of the Vessel Phantom with a Real Blood Vessel} 
\label{sec:ResultsPhantom}
The paragraph \emph{\nameref{sec:FabricationPhantom}} emphasized that a phantom that provides properties as close as possible to the real object is needed. This means that a model is required that not only represents similar mechanical properties, but also exhibits similar fluorescent staining.

In intravital microscopy studies of vascular systems, endothelial cells are often used to identify the blood vessels and membrane proteins specific for endothelial cells are fluorescently stained in order to obtain high contrast and to visualize the vascular lumen. 
Common intravital labeling methods are e.g.\ immunological staining with primary fluorescently labeled antibodies or primary-secondary antibody systems, or using the presence of other receptor molecules, such as acLDL \cite{Menon2015}, to bind customized fluorescent dye molecules \cite{Voyta1984Dec}. 
This, however, is not easily reproducible with a polymer-based tissue phantom, since it does not exhibit a uniquely functional surface structure or any other form of distinct binding domains for the labeling process.
Nevertheless, we were able to take advantage of another fundamental property of the tissue phantom: its hydrophobicity. 
The use of a water-soluble dye in combination with the hydrophobic PDMS results in a comparable staining pattern as that of endothelial cells as illustrated in \cref{fig:PhantomComparison}. 
Here, the direct comparison of 2P excited fluorescence images recorded from an immunofluorescntly labeled murine carotid artery and our custom-made PDMS tissue phantom is shown. The images from both experiments are remarkably similar as far as the shape is concerned.
We attribute this behavior to hydrophobic interactions, where the non-polar domains of the PDMS facilitate the adsorption of specific hydrophobic molecules. While this phenomenon is typically an undesirable effect in microfluidics due to the potential adsorption of contaminants, in this context it results in a distinctive staining pattern \cite{Wang_2012} similar to staining an endothelial layer in animal models.
However, due to absorption of the infrared femtosecond laser beam by water, as well as refractive index changes between the highly curved PDMS and the water inside the cavity, a significant loss of signal occurs near the bottom layer of the tissue phantom, which is quite notable in the cross-section in \cref{fig:PhantomComparison} (a).

\subsection{Performance of the Motion Correction Algorithm }
\begin{figure*}[ht]
	\centering
	\includegraphics[width=\textwidth]{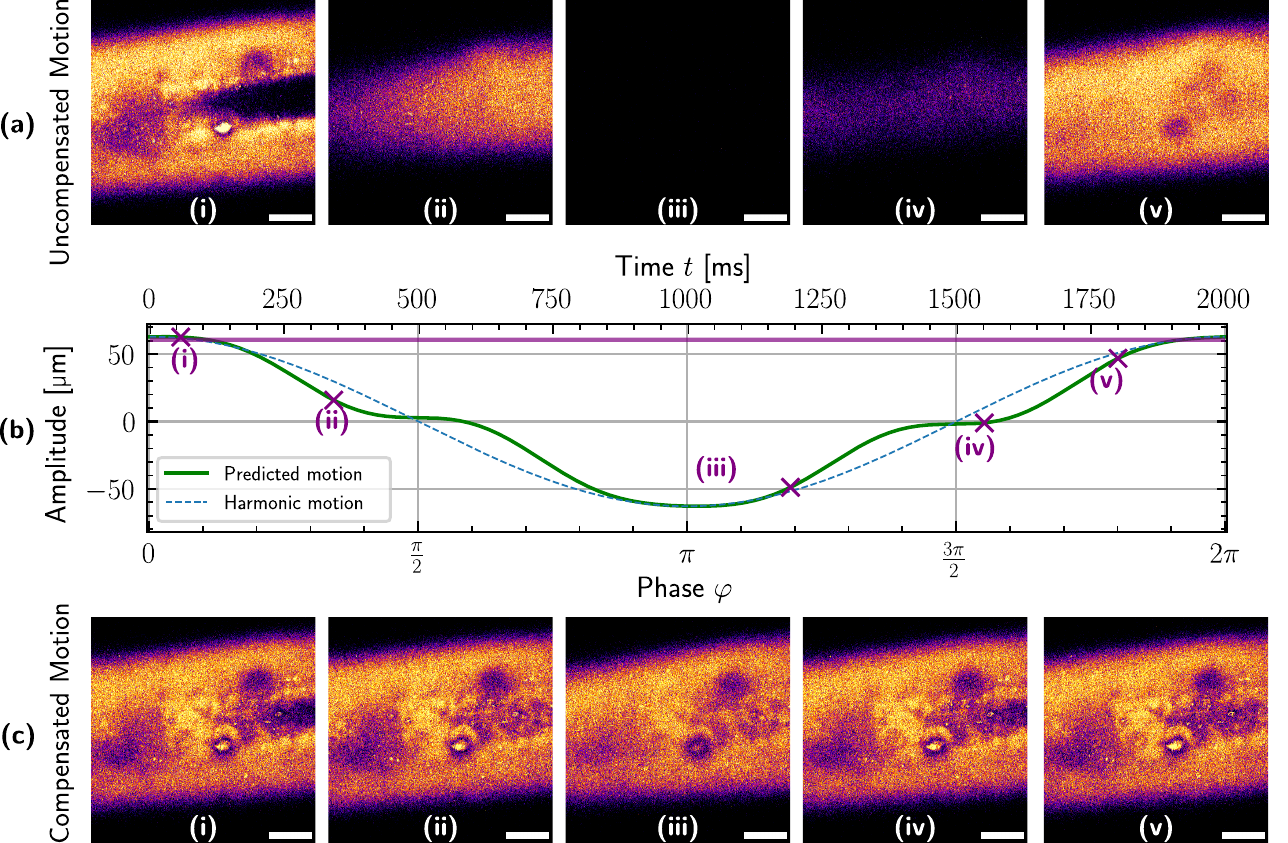}
	\caption{\textbf{Active motion compensation results obtained with the vessel phantom} -- (a) the uncompensated oscillation of the vessel phantom about the desired $z$-plane. The variation of the in-focus cross-sectional area of the cylinder is captured throughout an entire period. 
		The enumeration of (i) - (v) is supposed to indicate the selected points in time during the 0.5 Hz periodic motion interval.
		(b) illustrates the motion predicted by the algorithm (green curve) and the corresponding timestamps of the image acquisition from (a) or (c), respectively, with the purple markers. A positive amplitude means the specimen is approaching the objective lens. The values for the acquisition times are explicitly 
		(i):  $60 \text{ ms}$, (ii): $343 \text{ ms}$, (iii): $1\,194 \text{ ms}$, (iv): $1\,552 \text{ ms}$, (v): $1\,800 \text{ ms}$ based on a cycle.
		For the sake of simplicity and for better comparison with a harmonic motion, we have shifted the displayed phase $\varphi$ by a global phase $\pi$ compared to the definition of $\theta(t)$ and set the origin of the time axis $t(\varphi = 0) = 0$. 
		Images during the initialization phase reveal a total focal plane displacement of $\sim$ 125 µm. The desired observation plane is highlighted in the purplish area. 
		Non-harmonic portions at inflection points are evident when comparing the predicted amplitude (green curve) with a harmonic cosine function (blue dotted line) at $ t \simeq 750$ ms and $t \simeq 1\,250$ ms within an oscillation period.
		Scale bar: 75 µm}
	\label{fig:Results}
\end{figure*}
Next, we actuated the vessel phantom shown in \cref{fig:PhantomComparison} with an amplitude and frequency that simulates the breathing of a living animal. To achieve this, we used the voice coil platform and the frequency generator described in \cref{fig:Setup}. The fluorescently labeled vessel phantom was placed on a custom-made 3D-printed adapter that is attached to the voice coil to fit underneath the microscope, as shown in \cref{fig:Setup} (c). We used a 0.5 Hz periodic oscillation with an amplitude of 4 V provided by the frequency generator.

Following the discussion in section \emph{\nameref{sec:MathematicalModel}}, we performed the mandatory steps for the algorithm to initialize the data acquisition for a complete model of the tissue phantom (see the diagram labeled "\hyperref[paragraph:MotionAnalysisPhase]{Motion analysis phase}" in \cref{fig:MathematicalFlowChart}).
The function $\theta(t)$ used in the \hyperref[paragraph:MotionAnalysisPhase]{motion analysis phase} is $\theta(t) := \frac{2 \pi}{2\,000} (t \mod{ 2\,000 })$ since one period lasts $2\,000$ ms.
After approximately 1 min.\ of continuous image acquisition, $\Nz\NperZ = 3000$ images had been collected. The data set, as visualized in \cref{fig:MathematicalTimeLine}, contained $164\,418\,850$ elements, and subsequently the parameter calculation started. During this phase the laser beam was automatically attenuated with the EOM shown in \cref{fig:Setup} in order to reduce photobleaching and, in the case of an actual animal model, to avoid potentially harmful effects. 

Due to the physical RAM limitations of the computer, we used at most $\Nmax = 1.5\cdot10^7$ elements of the data set. Thus, less then $10\%$ of the total acquired data points contributed to the parameter calculation. 
Subsequently, as described in the flow chart in \cref{fig:MathematicalFlowChart} and the diagram labeled \hyperref[paragraph:MotionAnalysisPhase]{"Motion analysis phase"}, the coefficients of the \hyperref[paragraph:TruncatedFourierSeries]{truncated Fourier series} of $\msgMatrix{M}(\theta)$ and $\msgVector{b}(\theta)$ are fitted.
A visualization of the cylinders belonging to the parameters for different phases can be found in \cref{fig:MathematicalTimeLine} (c). 
The entire calculation to obtain the final coefficients took 2-5 min.\ - limited mostly by the time that it takes the computer to purge and allocate memory. 

Next, the mathematical model was invoked to track and subsequently stabilize the applied motion based on the recorded data. The results are presented in \cref{fig:Results}.
Besides various other parameters, the mathematical model is able to stabilize the amplitude of the total axial displacement we applied to the sample.
In \cref{fig:Results} (b) the resulting graph shows the predicted $z$-position at a certain phase point (resp. time point for a perfectly periodic oscillation). Therefore, the known electrical amplitude of 4 V applied to the voice coil corresponds to a mechanical motion of 125.8 µm in the axial direction. The frame rate during image acquisition was 30 fps with a FOV of 397 µm $\times$ 397 µm, acquiring 512 $\times$ 512 pixels for a single frame. 

Although the type of translational motion applied is not expected to induce deformations of the sample, the algorithm takes this possibility into account in order to generalize our procedure for future applications in more advanced and realistic simulations of animal motion.
Therefore, more parameters than just the position of the center, which is shown in \cref{fig:Results} (b) can be calculated. Other parameters that can, for example, be extracted from the results of the algorithm are the inclination of the vessel, which describes the rotation of the approximately cylindrical shape, and the variation of the diameter\footnote{In case of an elliptical cross-section we can obtain two diameters that equal twice the length of the major or minor axis} or the cross-section, which characterizes possible deformations such as expansions and contractions. 
The tilt of the principal axes determined from the tissue model does not show any significant variance, as the examined tilt of the main axes of the cylinder varied by less than $1\,^\circ$ over a full cycle. The predicted cross-sectional area varied by $\sim 17\,800 \text{ µm}^2$, which corresponds to a difference in circular diameter of $\delta D\sim 28 \text{ µm}$.
In addition, the algorithm allows us to extract the predicted motion of individual points of interest. The comparison of the data retrieved by the algorithm near the vessel center and the top movement showed only minor differences underlining the absence of any major deformation, which we attribute to the stiffness of the PDMS structure. 

In order to observe the effects of prolonged exposure of the PDMS channel to Nile blue, we reexamined the vessel phantom several weeks after the initial experiments. The absolute value of the diameter of the cylinder predicted by the algorithm\footnote{The metric for the measurement is the separation of the center points of the edges.} was now determined to be $\approx 400~\text{µm}$. We attribute this noticeably increased diameter observed in comparison to the results obtained in \cref{fig:PhantomComparison} to the fact that the prolonged exposure of the PDMS to Nile blue allowed the dye to diffuse further into the PDMS channel walls. This, in turn, resulted in an apparent increase in the diameter of the vessel phantom as extracted by the algorithm. Furthermore, the algorithm takes into account the outer diameter of the vessel borders when predicting its size, which also contributes to the observed apparent increase of the vessel diameter.

The primary goal of our current work was, however, to demonstrate our ability to actively compensate for the shift of the focal plane induced by axial motions comparable to those observed in respiratory animal models. This can be achieved through the rapid mechanical adjustment of the $z$-position of the objective lens, as facilitated by a piezoelectric transducer (PIFOC). The PIFOC itself has an intrinsic delay time of $\sim 65\ \text{ms}$, which is the time from when a command is sent to the controller to move to a $z$-position until the PIFOC has settled at this new position. It is operating under closed-loop control, and can cover a total $z$ distance of 400 µm. Thus, this system is able to, in principle, completely counteract the applied motion amplitude in real time. 
However, due to the calculations and commands sent to the microscope, an additional delay is induced. The theoretical frame rate limit is impacted most severely by the maximum operating speed of the PIFOC. 
Since the galvanometric scan mirrors are scanning the sample at a fixed (resonant) frequency, image acquisition can only be started at discrete points in time. This causes an additional waiting time and, combined with the time required for calculations and addressing the PIFOC, effectively reduces the frame rate for active motion compensation to a maximum value of $\sim 10~\text{fps}$. 

To verify whether the anticipated vertical motion of the sample can be stabilized well based on the parameters obtained by the mathematical model, we maintained the focal plane at $\varzPercent = 0.425$ of the changing vertical vessel extension above the moving vessel center, as described in the paragraph \hyperref[paragraph:MotionCompensationPhase]{motion compensation phase} (see \cref{fig:MathematicalTimeLine} (d)). Physically speaking, this corresponds to the upper edge of the vascular phantom.  
The result of this experiment can be seen in \cref{fig:Results}. 

The location of the plane of interest near the top edge of the vessel phantom border was chosen intentionally.  This position is particularly demanding to stabilize with an externally applied motion since small variations in the axial position $z$ of the sample immediately lead to comparatively large variations in the image data. A drift of the observation plane at the top will ultimately result in the phantom fading out of the FOV, whereas a displacement of the observation plane deeper inside the vascular phantom reveals two distinguishable lines, which correspond to the channel walls. 
This is illustrated in \cref{fig:Results} (a), where we show select images recorded during the uninhibited oscillation of the vessel phantom that are processed to initialize the motion compensation algorithm.  \FloatBarrier
The motion we applied to the phantom had a non-harmonic nature (see  \cref{fig:Results}). This is caused by the mechanical setup, which is, in essence, a 3D-printed lever glued to the voice coil, which holds the sample underneath the microscope objective lens. This non-harmonic motion further complicates the successful motion compensation, which becomes particularly apparent when comparing the motion fitted by the algorithm in \cref{fig:Results} (b) with a perfect cosine oscillation. These non-harmonicities in the data lead to large Fourier frequency components in the oscillation.
In contrast to the images in \cref{fig:Results} (a), the PIFOC  follows the commands provided by the predictions of the algorithm in \cref{fig:Results} (b) to compensate for the shift of the focal plane. This is illustrated in \cref{fig:Results} (c), where we show corresponding images at (approximately) the same phase points of the oscillatory motion, acquired 217 periods after \cref{fig:Results} (a), allowing for a direct comparison with the images where the vertical motion was not compensated. The deliberate selection of the different points shown here serves a dual purpose: first, it demonstrates the periodic nature of the oscillation through the image series in \cref{fig:Results} (a); second, these phase points exhibit distinct characteristics that are demanding for the algorithm and physical motion compensation.
The uncompensated $z$-positions of the sample at the acquisition times labeled (i) and (v) in \cref{fig:Results} closely align with the desired focal plane when the objective lens is in its equilibrium position. However, subtle variations in height are immediately discernible in the image data. Thus, the algorithm has to accurately predict and trace this motion with the help of the PIFOC .
The acquisition phase labeled (ii) depicts a location with a rapid local change in the plane of interest deep inside the PDMS structure. The comparison with the perfect harmonic cosine wave in \cref{fig:Results} (b) indicates a significant deviation from a perfect harmonic wave. Despite the high-frequency Fourier components caused by the non-harmonic oscillation, the active motion compensation algorithm was able to compensate for the displacement of the focal plane relative to the vessel phantom. This is especially underlined by \cref{fig:Results} (c) (ii), where the deviation from the harmonic motion and the current rate of change of the observation plane in the phantom is greatest (c.f. \cref{fig:Results} (b)).
However, the image series in \cref{fig:Results} (c) conveys a seemingly stationary vessel phantom, and hence illustrates the ability of the algorithm to overcome these challenges. No major visible displacements are present, and the focal plane precisely tracks the top edge of the cylindrical vessel phantom. The model, therefore, stabilized the $z$-position correctly at the respective point in time. 
Further challenges arise from the remaining acquisition times depicted in \cref{fig:Results}, as the acquisition phase (iii) is the furthest away from our actual plane of interest. Finally, the non-harmonic part of the externally applied oscillation at (iv) requires accurate prediction of the motion, particularly the non-harmonic nature, in order to maintain the observation plane at the top of the phantom. 

The images in (c) (i)-(v) thus demonstrate the consistency of the predictions made by the algorithm with the actual motion, despite the additional challenges posed by the non-harmonic nature of the oscillation.

Only minor deviations are apparent in the compensated images in \cref{fig:Results} (c), which could potentially be caused by the model's uncertainty in time estimation of up to $\frac{1}{30}$ s caused by the finite image acquisition or scan time, respectively, and the unknown exact time when the scan starts.

This becomes evident in \cref{fig:Results} (c) (i) and (iv). The axial position of the observation plane lies slightly below the surface of the sample, which is demonstrated by a gap that is perceivable between the channel walls. In consideration of the current oscillation phase for the frame acquisition, which we can extract from \cref{fig:Results} (b), we recognize that the $z$-position of the sample was slightly lower for each of the images (i) and (iv). Equivalently, the observation plane was shifted closer to the top of the vessel phantom. 
Thus, the incorrect prediction of the $z$-position could be due to the time difference between the actual and the predicted image acquisition time.
This implies that we could improve the model further by taking into account the possible image acquisition times and selecting the future time point $t$ from the set of possible image acquisition times. 

\section{Conclusion}
We successfully compensated the motion of a vessel phantom in 2P-fluorescence microscopy. We were able to stabilize the focal plane inside the vessel phantom as depicted in \cref{fig:Results} for several minutes amid simulated external respiratory motion.
This underscores the fundamental capability of our setup and algorithm that will be leveraged in future motion compensation experiments with living specimens. This should enable us to observe inflammatory processes in vivo in real-time.
We have demonstrated a novel preparation method resulting in stained vessel phantoms that provide sufficient similarities to a murine carotis communis. In order to further boost the rates at which our active motion compensation can operate, we plan to replace the PIFOC with an adaptive optics module. The latter can shift the focus position more rapidly in the $z$- but also in the $xy$-plane, which should then enables the compensation of the complex motion observed in living specimens. In particular we hope to accomplish pixel- or line-wise motion correction during the scanning process of a single image, which is required when examining more rapid motions with larger amplitudes. Moreover, we plan to study the effects of tissue deformation due to peristaltic motion and options to detect and compensate these effects with the help our mathematical algorithm.

Considering improvements to the model:
in practice, animal motion is more complex than just the movement along a single dimension. The mathematical model that was used here can be extended to analyze and compensate for $n$ motions with different frequencies. 
For this purpose, we model $\theta(t) \in [0, 2 \pi)^n$, where the $i$th component describes the phase of the $i$th motion.
Instead of using a \hyperref[paragraph:TruncatedFourierSeries]{truncated Fourier series}, we perform a truncated multidimensional Fourier series. 
The model can also be applied when the motion frequency is not constant but varies,
in which case the current frequency and phase could be estimated from the imaging data.

In the current implementation it is ignored that the pixels of an image are actually acquired at different time points.
Taking, in both motion analysis and correction phase, the exact acquisition time of each single pixel into account
(which can readily be computed from the starting time of the image acquisition and the scanning trajectory)
will improve the quality of the extracted motion and allow motion compensation even during the image acquisition.

\begin{acknowledgement}
The authors thank the Deutsche Forschungsgemeinschaft (DFG, German Research Foundation)) for funding this project through the CRC 1450 – 431460824 and through Germany's Excellence Strategy EXC 2044 –390685587, Mathematics Münster: Dynamics–Geometry–Structure. TH and FK gratefully acknowledge DFG grants INST 215/614-1 FUGG and INST 211/900-1 FUGG for the purchase of their multiphoton microscopes. We also thank Jasmin Schürstedt for her help with PDMS sample preparation.

\end{acknowledgement}

\bibliography{bibliography_acs_photonics}
\end{document}